\documentclass[twocolumn,preprintnumbers,amsmath,amssymb]{revtex4}
\usepackage{graphicx} 
\usepackage{setspace} 
\usepackage{dcolumn} 
\usepackage{bm} 
\usepackage{natbib}
\usepackage{hyperref}
\hypersetup{colorlinks=true,linkcolor=blue,citecolor=blue,urlcolor=blue}
\usepackage{epsfig}
\begin{document}
\title{Novel entropy-stabilized fluorite oxides with multifunctional properties}
\author{Ashutosh Kumar}
\author{David Bérardan}
\author{Francois Brisset}
\author{Diana Dragoe}
\author{Nita Dragoe\footnote{Email: nita.dragoe@universite-paris-saclay.fr}}

\affiliation{ICMMO (UMR CNRS 8182), Université Paris-Saclay, F-91405 Orsay, France}
\date{\today}
\begin{abstract}
Development of new high-entropy oxides having configurational entropy dominating the phase stability has become a hot topic since the discovery of rock salt structure entropy-stabilized (MgCoNiCuZn)O in 2015. Herein, we report a set of novel entropy-stabilized fluorite oxides: Zr$_{0.2}$Hf$_{0.2}$Ce$_{0.2}$Sn$_{0.2}$Mn$_{0.2}$O$_{2-\delta}$ Zr$_{0.2}$Hf$_{0.2}$Ti$_{0.2}$Mn$_{0.2}$Ce$_{0.2}$O$_{2-\delta}$, Zr$_{0.225}$Hf$_{0.225}$Ti$_{0.225}$Mn$_{0.225}$Ce$_{0.1}$O$_{2-\delta}$, and Zr$_{0.2}$Hf$_{0.2}$Ti$_{0.2}$Mn$_{0.2}$Ce$_{0.1}$Ta$_{0.05}$Fe$_{0.05}$O$_{2-\delta}$ synthesized using standard solid-state reaction. These compounds have been investigated using X-ray diffraction, scanning electron microscopy, and X-ray photoelectron spectroscopy techniques to discern their structural, microstructural, and chemical properties. The configurational-entropy dominated phase stability and hence the entropy stabilization of the compounds is confirmed by cyclic heat treatments. The mismatch in the ionic radii and oxidation state of the cations are the key factors in achieving a single-phase fluorite structure.  Further, screening of physical properties including thermal conductivity, optical band gap, magnetic properties, and impedance spectroscopy is discussed. Thermal conductivity of 1.4--1.7 W·m$^{-1}$·K$^{-1}$ is observed at 300 K and remains mostly invariant across a wide temperature range (300\,K--1073\,K), favorable for thermal barrier coating applications. These entropy-stabilized samples have an optical band gap of 1.6--1.8 eV, enabling light absorption across the visible spectrum and hence could be promising for photocatalytic applications. The impedance spectroscopy data of the entropy-stabilized samples reveal the presence of electronic contributions with small activation energy (0.3--0.4 eV) across a temperature range of 298K-423K. These observations in entropy-stabilized fluorite systems show potential for their multifunctional applications via further optimization and confirm the great chemical versatility of entropy-stabilized oxides.
\end{abstract}
\maketitle
\section{Introduction}
Discovery of novel materials with enhanced functional properties is besought by the increasing dependence of humankind on technology. Among the many recent discoveries, entropy-stabilized oxides (ESO) are a new class of ceramic materials where, in general, five or more cations in the equimolar ratio are integrated at a single crystallographic site.  to achieve entropic dominance to Gibb's free energy that drives the thermodynamic stability.\cite{musico2020emergent,dragoe2019order} These materials were discovered when Rost et. al.\cite{rost2015entropy} extended the high-entropy alloys concept (Cantor et. al.,\cite{cantor2004microstructural} Yeh et al.,\cite{yeh2004nanostructured}) to ionic bonded ceramic and synthesized the first entropy-stabilized oxide, (Mg$_{0.2}$Co$_{0.2}$Ni$_{0.2}$Cu$_{0.2}$Zn$_{0.2}$)O, with rock-salt structure, and drove an impetus in material science research based on ceramics. It was followed by the findings of new high-entropy materials with different crystal structures which include fluorite,\cite{chen2018five,djenadic2017multicomponent} spinels,\cite{mao2020novel,chen2020new,musico2019tunable} bixbyite,\cite{sarkar2017multicomponent} perovskite,\cite{jiang2018new} rutile,\cite{kirnbauer2019high} pyrochlore\cite{karthick2021phase, teng2020synthesis, vayer2021new}, etc. These high-entropy oxides show interesting physical properties such as large lithium-ion conductivity,\cite{berardan2016room} ultralow thermal conductivity,\cite{li2019high} improved figure of merit for thermoelectric,\cite{kumar2023thermoelectric}, exotic magnetic ordering,\cite{musico2019tunable,kumar2023magnetic} colossal dielectric constant,\cite{berardan2016colossal} proton conductivity,\cite{gazda2020novel} photocatalysis,\cite{sun2021high, edalati2020photocatalytic} etc. Therefore, they constitute a very exciting playground for the development of new functional oxides.\\ 
It should be noted that the terms high-entropy and entropy-stabilized oxides have often been used interchangeably in the literature although they refer to two different concepts.\cite{musico2020emergent, dragoe2019order, sarkar2020high} Populating a specific site with multiple cations results in high configurational entropy, however, for the system to be entropy-stabilized, it should possess both high configurational entropy along with positive enthalpy of mixing, indicating that the thermodynamic stability is driven by entropy.\cite{oses2020high} It is known that the Gibbs free energy of mixing ($\Delta$G$_{mix}$) should reduce in order to form a single phase and is influenced by the change in enthalpy of mixing ($\Delta$H$_{mix}$) and change in entropy of mixing ($\Delta$S$_{mix}$) as $\Delta$G$_{mix}$=$\Delta$H$_{mix}$-T$\Delta$S$_{mix}$. It is evident that for the case with $\Delta$H$_{mix}$ $<$ 0 and $\Delta$H$_{mix}$=0 (ideal solution), $\Delta$G$_{mix}$ is negative, which favors the formation of single-phase solid solution, except when $\Delta$H$_{mix}$ $<<$ 0 that leads to the formation of line compounds.\cite{murty2019high} Oppositely, a positive $\Delta$H$_{mix}$ $>$ 0, for example, due to large differences in atomic sizes or electronegativity, generally precludes the formation of single-phase compounds. In that case, high configurational entropy is required to dominate the free energy and overcome the positive $\Delta$H$_{mix}$. The stability of entropy-stabilized phases has a dependence on the decreased/increased contribution of configurational entropy at low/high temperatures (T$\Delta$S$_{mix}$). Thus, a reversible phase transformation of low-temperature (enthalpy favorable) multiple phases to a high-temperature (entropy-stabilized) single-phase compound constitutes a signature for entropy-stabilization. This phase transition is endothermic in nature when screened using differential scanning calorimetric analysis.\cite{rost2015entropy} Therefore entropy-stabilization of a compound can be tested via cyclic heat treatment coupled with X-ray diffraction, differential scanning calorimetry, scanning electron microscopy, etc. For example, several recent studies including (Ce$_{0.2}$La$_{0.2}$Pr$_{0.2}$Sm$_{0.2}$Y$_{0.2}$)O$_{2-\delta}$,\cite{sarkar2020role} (LaNdPrSmEu)$_{1-x}$Sr$_x$(Co/Mn)O$_3$,\cite{kumar2023thermoelectric, kumar2023magnetic} (Co$_{0.2}$Cr$_{0.2}$Fe$_{0.2}$Mn$_{0.2}$Ni$_{0.2}$)$_3$O$_4$,\cite{sarkar2022comprehensive} etc. do not show reversible phase transformation and hence correspond to high-entropy systems but not entropy-stabilized ones. The upholding of the single-phase in these systems may be attributed to the favorable enthalpy of mixing ($\Delta$H$_{mix}$ $<$ 0) with minimum influence of configurational entropy (T$\Delta$S$_{mix}$). Some empirical selection rules have been suggested to synthesize entropy-stabilized oxides that include: the difference between the ionic radii of the different cations at the single crystallographic site should be small, one of the cations should have a different usual local environment (for example in the case of (Mg$_{0.2}$Co$_{0.2}$Ni$_{0.2}$Cu$_{0.2}$Zn$_{0.2}$)O MgO, CoO and NiO crystallize in a rocksalt structure whereas CuO and ZnO crystallize respectively in tenorite and wurtzite structures), the overall system should have distinctive electronegativity, etc., which are explained in details elsewhere.\cite{musico2020emergent, rost2015entropy, djenadic2017multicomponent, sarkar2018rare, mccormack2021thermodynamics} However, only a small ionic radii mismatch is considered for designing high entropy oxides.\\
In 2018, Chen et al. reported the first entropy-stabilized fluorite, Zr$_{0.2}$Hf$_{0.2}$Ce$_{0.2}$Ti$_{0.2}$Sn$_{0.2}$O$_2$ with thermal conductivity ($\kappa$) $\sim$ 1.28 W·m$^{-1}$·K$^{-1}$ at 300\,K.\cite{chen2018five} Further, Gild et al.,\cite{gild2018high} showed a series of high-entropy fluorite oxides, followed by synthesis of (Zr$_{0.2}$Ce$_{0.2}$Hf$_{0.2}$Y$_{0.2}$Al$_{0.2}$)O$_{2-\delta}$ by Wen et al.,\cite{wen2022processing} However, entropy-stabilization was not established in these systems (cyclic heat treatment or other methods were not shown). Ding et. al.,\cite{ding2022hf0} and He et. al.,\cite{he2021four} reported (Hf$_{0.25}$Zr$_{0.25}$Sn$_{0.25}$Ti$_{0.25}$)O$_2$ system with orthorhombic structure (space group: \textit{Pbcn}); however, no evidence was provided to support their claim of entropy stabilization. The high-entropy fluorite oxides also include reports by Djenadic et al on equiatomic rare-earth oxides having high configurational entropy (Ce$_{0.2}$La$_{0.2}$Sm$_{0.2}$Pr$_{0.2}$Y$_{0.2}$)O$_{2-\delta}$\cite{djenadic2017multicomponent} followed by (Nd$_{0.16}$Ce$_{0.16}$La$_{0.16}$Sm$_{0.16}$Pr$_{0.16}$Y$_{0.16}$)O$_{2-\delta}$ by Sarkar et al.,\cite{sarkar2017multicomponent} Recently Kante et al, fabricated (CeLaSmPrY)O$_{2-x}$ fluorite thin film using sol-gel and pulsed laser deposition processes and shows their optical band gap.\cite{kante2022high} Also, Chen et al, showed a series of samples where each element in Zr$_{0.2}$Hf$_{0.2}$Ce$_{0.2}$Ti$_{0.2}$Sn$_{0.2}$O$_2$ is replaced by Ca to further enlarge the fluorite family and concluded that Ce is important to realize single-phase high-entropy fluorite oxide.\cite{chen2023formation} In this article, we explore a set of novel entropy-stabilized fluorite oxides and in this process, we also reinvestigate the first reported high-entropy fluorite oxides by Chen et. al.,\cite{chen2018five} and observe a peculiar structural behavior on the surface and bulk in Zr$_{0.2}$Hf$_{0.2}$Ce$_{0.2}$Ti$_{0.2}$Sn$_{0.2}$O$_2$, attributed to the possible reduction of Ce$^{4+}$ for this combination of elements. Further, a deeper understanding of single-phase formation in high-entropy fluorite oxides is developed, and novel entropy-stabilized fluorite oxides: Zr$_{0.2}$Hf$_{0.2}$Ce$_{0.2}$Sn$_{0.2}$Mn$_{0.2}$O$_{2-\delta}$, Zr$_{0.2}$Hf$_{0.2}$Ti$_{0.2}$Mn$_{0.2}$Ce$_{0.2}$O$_{2-\delta}$, Zr$_{0.225}$Hf$_{0.225}$Ti$_{0.225}$Mn$_{0.225}$Ce$_{0.1}$O$_{2-\delta}$, Zr$_{0.2}$Hf$_{0.2}$Ti$_{0.2}$Mn$_{0.2}$Ce$_{0.1}$Ta$_{0.05}$Fe$_{0.05}$O$_{2-\delta}$ synthesized using standard solid-state reaction route are reported for the first time. These novel high-entropy fluorite samples are further screened using cyclic thermal treatment to establish the configurational entropy-dominated phase stability which is further supported by scanning electron microscopy-energy dispersive X-ray spectroscopy (SEM-EDS) techniques. Last, preliminary studies of their physical properties including optical band gap, thermal conductivity ($\kappa$), magnetic properties (M--H and M--T), low-temperature specific heat capacity, and impedance spectroscopy (IS) studies are discussed.\\

\section{Experimental Section}
\begin{figure}
\centering
  \includegraphics[width=0.85\linewidth]{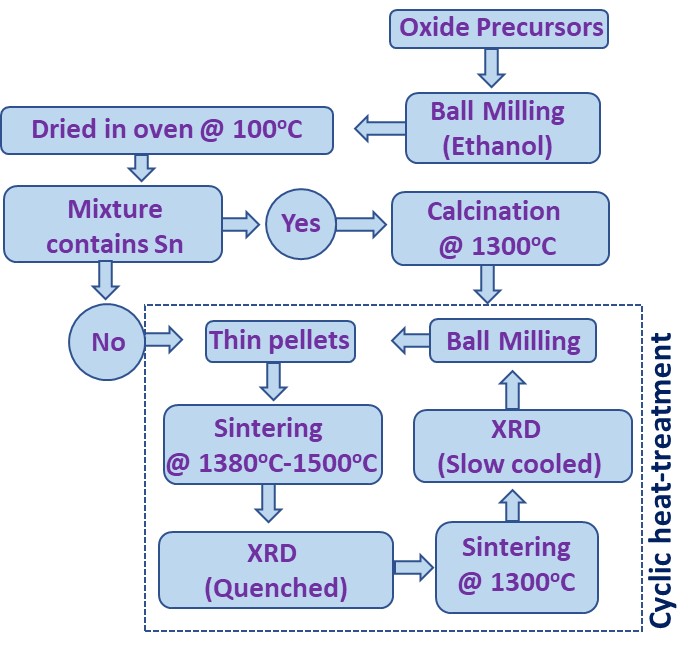}
  \caption{Schematic presentation of the synthesis cycle used for the preparation of high-entropy fluorite oxides.}
  \label{synthesis}
\end{figure}

High-entropy fluorite samples were synthesized using standard solid-state reaction methods. The details of oxide precursors used for each synthesis are shown in Table S1. Stoichiometric amounts of oxide precursors for each composition were mixed using planetary ball milling (Fritsch Pulverisette 7 Premium Line) at 350 rpm for 60 cycles (5 min on and 5 min off time) in a wet medium (ethanol). The slurry obtained after ball milling was dried overnight inside an oven at 100$^{\circ}$C. The Sn-containing samples were calcined at 1300$^{\circ}$C for 6 hours with a heating and cooling rate of 200$^{\circ}$C/hour. The obtained powder was again ball milled and compressed into cylindrical pellets of 10 mm in diameter and 1.5--2 mm thickness. However, this intermediate calcination step was skipped for samples without Sn. The compacted pellet was placed on a platinum (Pt) sheet inside an alumina boat and sintered in the temperature range from 1380$^{\circ}$C--1500$^{\circ}$C for 6--15 hours followed by fast quenching. Cyclic heat treatments were performed to establish the entropy-stabilized phase formation of the samples. The scheme for the synthesis and entropy-stabilization test is shown in Fig.~\ref{synthesis}. The exact temperature of sintering used for each sample is shown in Table~\ref{table1}.\\
\begin{figure*}
\centering
  \includegraphics[width=0.99\linewidth]{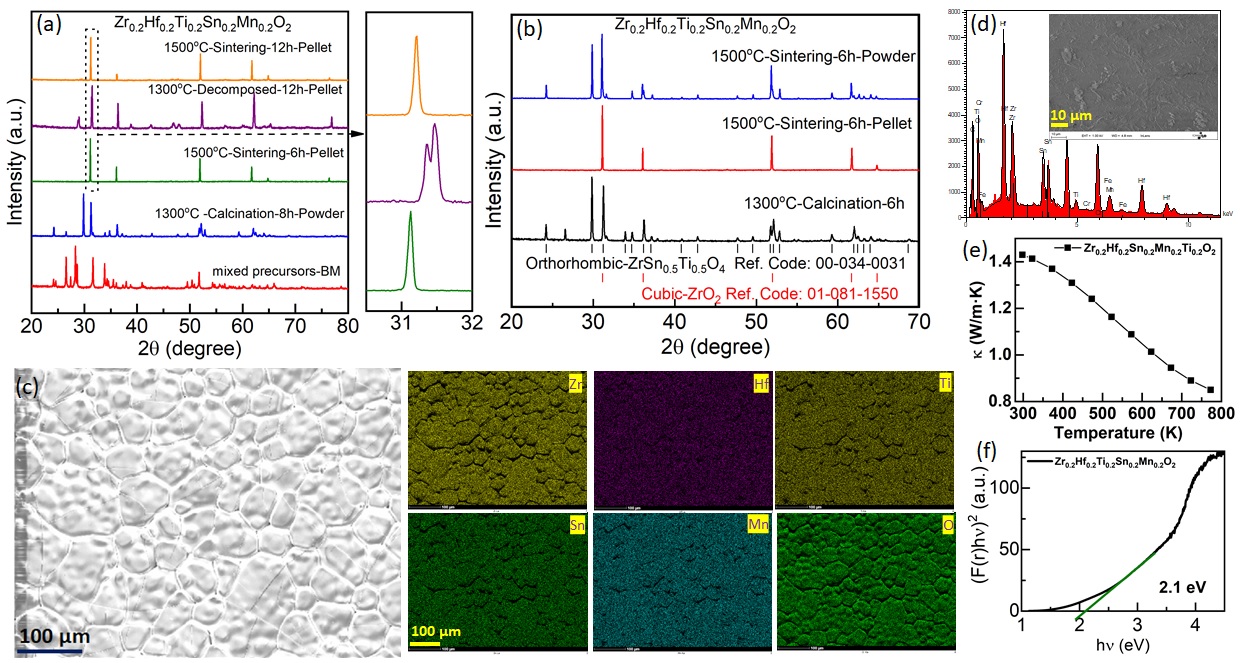}
  \caption{(a) X-ray diffraction pattern of Zr$_{0.2}$Hf$_{0.2}$Ti$_{0.2}$Sn$_{0.2}$Mn$_{0.2}$O$_{2-\delta}$ at different synthesis steps (mixed precursors using ball milling (BM), calcination, and final sintering) followed by entropy-stabilization test. (b) XRD patterns obtained on the surface (Pellet) and bulk (powder) sample sintered at 1500$^{\circ}$C and its comparison with the XRD pattern of the same sample calcined at 1300$^{\circ}$C. (c) scanning electron microscopy (SEM) image and corresponding elemental mapping of each element using energy dispersive X-ray spectroscopy (EDS) (d) corresponding EDS spectra, Inset: SEM image for Zr$_{0.2}$Hf$_{0.2}$Ti$_{0.2}$Sn$_{0.2}$Mn$_{0.2}$O$_{2-\delta}$. (e) Thermal conductivity ($\kappa$) and (f) band gap obtained from UV-Visible spectroscopy for Zr$_{0.2}$Hf$_{0.2}$Ti$_{0.2}$Sn$_{0.2}$Mn$_{0.2}$O$_{2-\delta}$.}
  \label{Fig2}
\end{figure*}
The structural characterization of the samples was performed by X-ray diffraction using a Panalytical X'Pert diffractometer with a Ge(111) incident monochromator, a copper tube (K$_{\alpha1}$ radiation), and a fast detector (X'celerator). Surface morphology and elemental mapping were observed using scanning electron microscopy (SEM-FEG Zeiss Sigma HD) equipped with energy-dispersive X-ray spectroscopy (SAMx-IDFix EDS) operating at 20 kV. Further, X-ray photoelectron spectroscopy (XPS) measurements were performed on a Thermo Fisher Scientific instrument with a monochromatic Al-K$_{\alpha}$ X-ray source (energy 1486.68 eV) and a hemispherical analyzer. The base pressure was around 5$\times$10$^{-9}$ mbar and the diameter of the X-ray beam spot was 400 $\mu$m, corresponding to an irradiated surface of approximately 1 mm$^2$. The hemispherical analyzer was operated at 0$^{\circ}$ take-off angle in the Constant Analyzer Energy (CAE) mode. Wide scan spectra were recorded at pass energy of 200 eV and an energy step of 1 eV while narrow scan spectra were recorded at pass energy of 50 eV and an energy step of 0.1 eV. Charge compensation was achieved by employing a "dual beam" flood gun, using low-energy electrons ($<$5 eV) and argon ions. The binding energy scale was calibrated on the neutral carbon set at 285 eV. The obtained core-level spectra were treated using CasaXPS software.21 The thermal conductivity ($\kappa$) of the samples was calculated using $\kappa$=D·$\rho_s$·C$_p$. The thermal diffusivity (D) was measured using the laser flash technique (Netzsch LFA 427), and sample density ($\rho_s$) was calculated using the mass of the pellet and its geometric volume. The specific heat capacity (C$_p$) was calculated using the Dulong-Petit law. The optical band gap was obtained from UV-Visible spectroscopy measurement using Agilent Cary 5000 UV-Vis-NIR spectrometer in diffuse reflective spectra (DRS) mode. The obtained reflectance data is converted into absorbance spectra using the Kubelka-Munk function. The magnetic measurements (M--H at 2 K and M--T in ZFC and FC modes at 100\,Oe) were obtained using a SQUID magnetometer, (MPMS, Quantum Design, USA). The electrical properties of entropy-stabilized samples were investigated using the Impedance spectroscopy (IS) technique. The IS measurements were carried out on cylindrical samples coated with platinum on both sides (Pt/HEOx/Pt, using DC magnetron sputtering) at different temperatures in the frequency range of 0.01--3.3$\times$10$^6$ Hz and 300 mV disturbance amplitude. The samples were placed in a Swagelok-type casing that was then placed inside an oven for temperature-dependent measurements. The obtained IS data were analyzed using fit3Zarcs software.\cite{dragoezarc3fit} Electrical resistivity was also measured in the two-probe configuration using a Keithley source meter (Keithley 6517B) from 298\,K--423\,K. Low-temperature specific heat measurements were carried out using a PPMS (Physical Property Measurement System), Quantum Design across a temperature range of 2K--300K. The sample was kept in a vacuum of circa 0.01 mTorr and protected by an anti-radiation shield, to prevent dissipation from convection and thermal radiation respectively. An Apiezon N grease was used to ensure heat transfer from the sample to the sample holder.\\
\section{Results and Discussion} 
\begin{figure*}
\centering
  \includegraphics[width=0.99\linewidth]{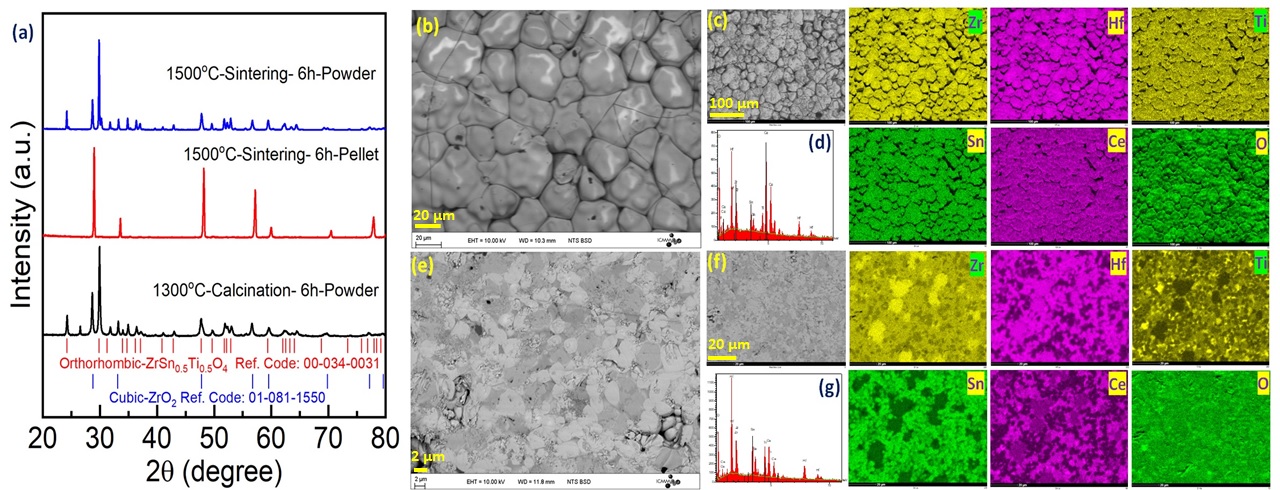}
  \caption{(a) X-ray diffraction pattern for Zr$_{0.2}$Hf$_{0.2}$Ce$_{0.2}$Sn$_{0.2}$Ti$_{0.2}$O$_2$ at each step. The XRD pattern on the pellet (surface) and powder (bulk) is shown. The Bragg's position corresponding to the orthorhombic and cubic phases is marked. Scanning electron microscopy images for Zr$_{0.2}$Hf$_{0.2}$Ce$_{0.2}$Sn$_{0.2}$Ti$_{0.2}$O$_2$ observed in backscattered electron (BSE) mode (b) on the surface (e) bulk is shown. The elemental mapping and energy dispersive x-ray spectra (EDS) on the (c-d) surface and (f-g) bulk is shown.}
  \label{fig3}
\end{figure*}

First, the outcome of a synthesis of the new composition Zr$_{0.2}$Hf$_{0.2}$Ti$_{0.2}$Sn$_{0.2}$Mn$_{0.2}$O$_{2-\delta}$ is shown in Fig.~\ref{Fig2}(a) at each step of the synthesis. The XRD pattern of the ball-milled (BM) oxide precursors consists of peaks corresponding to each binary oxide used for the synthesis. Further calcination of mixed precursors at 1300$^{\circ}$C results in a multiphase compound, with a main orthorhombic phase (space group: \textit{Pbcn}).\cite{ding2022hf0} After consolidating the powder in thin pellets and sintering at 1500$^{\circ}$C for 6 hours followed by quenching, the XRD pattern of the surface of the pellet (penetration depth of the XRD beam estimated to be a few µm) evidences a single-phase fluorite structure with no impurity within the sensitivity of the XRD. In order to assess the role of configurational entropy in the formation of the compound, this pellet was further exposed to cyclic heat treatment. After heating at 1300$^{\circ}$C for 12 hours, the XRD pattern of the pellet shows the presence of multiple phases, which are reversed back to a single-phase fluorite structure with further sintering at 1500$^{\circ}$C followed by quenching. The zoom-in image of Fig.~\ref{Fig2}(a) shows the evolution of the main peak of the XRD pattern of the pellet during the different steps of thermal cycling, demonstrating entropy stabilization. At a first glance, this observation confirms that the Zr$_{0.2}$Hf$_{0.2}$Ti$_{0.2}$Sn$_{0.2}$Mn$_{0.2}$O$_{2-\delta}$ phase may be entropy-stabilized. However, the powder obtained after grinding the single-phase pellet results in a very different and unexpected picture, as shown in Fig.~\ref{Fig2}(b). In this figure, a comparison of the XRD measurements done on the calcined precursors at 1300$^{\circ}$C (bottom), the surface of the pellet sintered at 1500$^{\circ}$C (middle), and grinded powder (bulk, top) is shown. It is seen that the XRD patterns of the surface (pellet) and bulk (powder) are very different. Although the XRD pattern of the pellet evidences a single-phase fluorite structure with the \textit{Fm-3m} space group, the powder consists of a combination of fluorite and orthorhombic structures.\cite{ding2022hf0} At this stage, the origin of such a distinctive nature of XRD patterns on the surface and bulk is not very clear. Also, such observation has never been reported in the literature for these high entropy compounds, to the best of our knowledge.\\
The energy dispersive X-ray spectroscopy (EDS) mapping performed on the surface of the pellet shows the homogeneous nature of the sample and further confirms that there is no impurity present on the surface (Fig.~\ref{Fig2}(c)). The EDS spectrum obtained from the surface of Zr$_{0.2}$Hf$_{0.2}$Ti$_{0.2}$Sn$_{0.2}$Mn$_{0.2}$O$_{2-\delta}$ is shown in Fig.~\ref{Fig2}(d). The atomic percentage obtained is in good agreement with the stoichiometric amounts of elements present in the system. Besides, the thermal conductivity at 300K is 1.42 W·m$^{-1}$·K$^{-1}$ (Fig.~\ref{Fig2}(e)) and is in close agreement with the value (1.28 W·m$^{-1}$·K$^{-1}$) reported by Chen et. al. for Zr$_{0.2}$Hf$_{0.2}$Ce$_{0.2}$Sn$_{0.2}$Ti$_{0.2}$O$_2$.\cite{chen2018five} The thermal conductivity decreases rapidly with a rise in temperature, however, the exact reason for such a change is difficult to understand in a multiphase sample. The optical band gap obtained from the UV-Visible spectroscopy for this sample is 2.1 eV (Fig.~\ref{Fig2}(f)).\\
Following this unexpected observation, we decided to reinvestigate the synthesis of the Zr$_{0.2}$Hf$_{0.2}$Ce$_{0.2}$Sn$_{0.2}$Ti$_{0.2}$O$_2$ compound following the procedure used by Chen et al.6 The XRD patterns corresponding to each step are shown in Fig.~\ref{fig3}(a). Similar to our new Zr$_{0.2}$Hf$_{0.2}$Ti$_{0.2}$Sn$_{0.2}$Mn$_{0.2}$O$_{2-\delta}$ composition, they also show a single-phase fluorite structure on the pellet (surface), whereas the XRD pattern of the grinded powder (bulk) corresponds to a multiphase sample with a predominant orthorhombic structure. This observation of a single-phase pattern on the surface (pellet) is consistent with the report by Chen et. al.\cite{chen2018five} However, the nature of the samples that were analyzed by these authors using XRD was not mentioned in their report. Besides, the (mostly) orthorhombic nature of the XRD pattern for the powder samples agrees with the recent studies for four-component binary high-entropy oxides using similar elements.\cite{ding2022hf0, he2021four} It is seen that the XRD pattern for the bulk is similar to the XRD pattern of the precursors calcined at 1300$^{\circ}$C, indicating that such mixture could be difficult to convert back to single-phase fluorite at 1500$^{\circ}$C due to kinetic issues. In order to avoid this possible difficulty, a new synthesis is done where the oxide precursors are mixed using ball milling, the mixture is consolidated into pellets and is directly sintered at 1500$^{\circ}$C followed by quenching. However, the XRD pattern (Fig.~S1) again shows the distinctive nature of the diffraction pattern in pellet and powder. Fig.~S2(a) shows the XRD pattern corresponding to each step during the synthesis of Zr$_{0.2}$Hf$_{0.2}$Ce$_{0.2}$Sn$_{0.2}$Ti$_{0.2}$O$_2$. Further, two more similar compositions Zr$_{0.2}$Hf$_{0.2}$Ti$_{0.2}$Sn$_{0.2}$Ce$_{0.1}$Y$_{0.1}$O$_{2-\delta}$ and Zr$_{0.2}$Hf$_{0.2}$Ti$_{0.2}$Sn$_{0.2}$Y$_{0.2}$O$_{2-\delta}$ have been synthesized and the XRD diffraction pattern also depicts the different XRD pattern in surface and bulk and is shown in Fig.~S2(b-c). A similar synthesis route has been followed for these two samples and a similar nature of diffraction pattern has been observed. However, it should be noted that when one of the elements has an oxidation state lower than 4+, (in these cases: Y$^{3+}$, or Mn$^{2+}$ (shown later in XPS)), the Bragg peaks of the fluorite phase appear significantly in the bulk phase as well, whereas they are hardly observed in Zr$_{0.2}$Hf$_{0.2}$Ce$_{0.2}$Sn$_{0.2}$Ti$_{0.2}$O$_2$. However, even in these cases, the samples still do not exhibit a single-phase fluorite structure. Several other compositions with Sn as one of the elements have been tried but no single-phase fluorite structure has been seen in the powder samples.\\
\begin{figure}
\centering
  \includegraphics[width=0.99\linewidth]{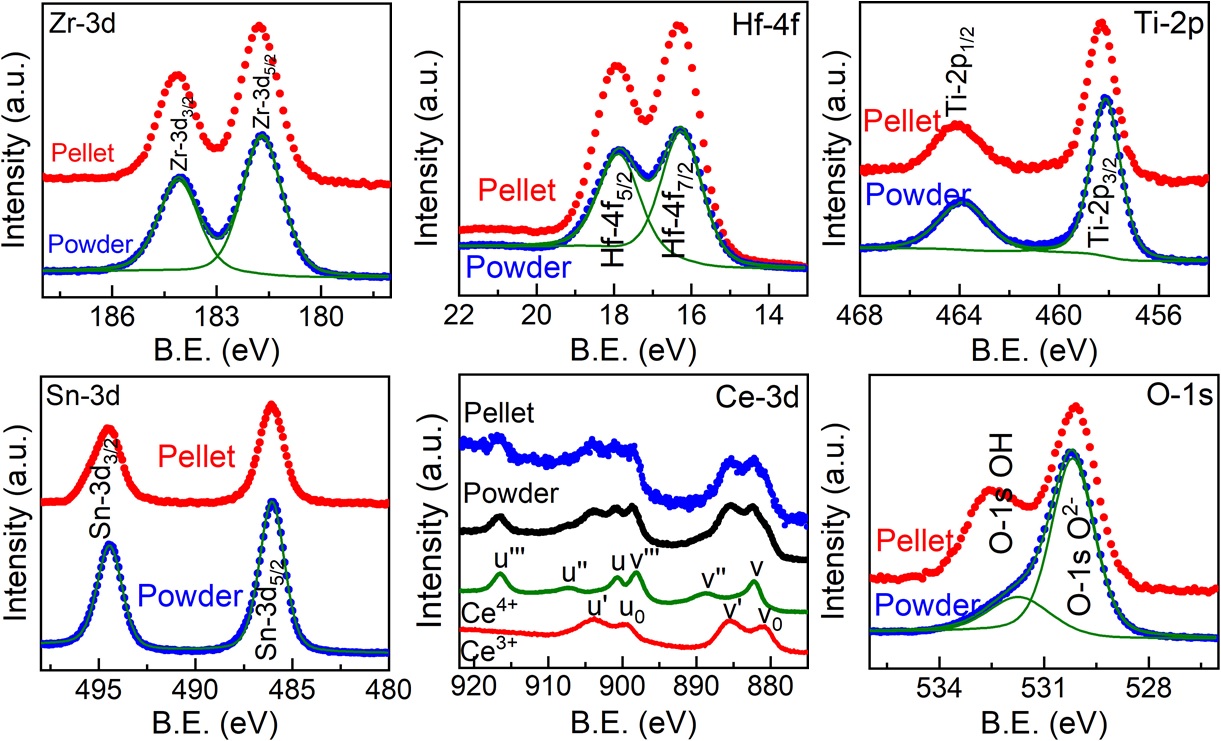}
  \caption{Core--level X-ray photoelectron spectroscopy (XPS) measurement carried on the surface and bulk for each element in Zr$_{0.2}$Hf$_{0.2}$Ti$_{0.2}$Sn$_{0.2}$Ce$_{0.2}$O$_2$.}
  \label{fig4}
\end{figure}
\begin{figure*}
\centering
  \includegraphics[width=0.99\linewidth]{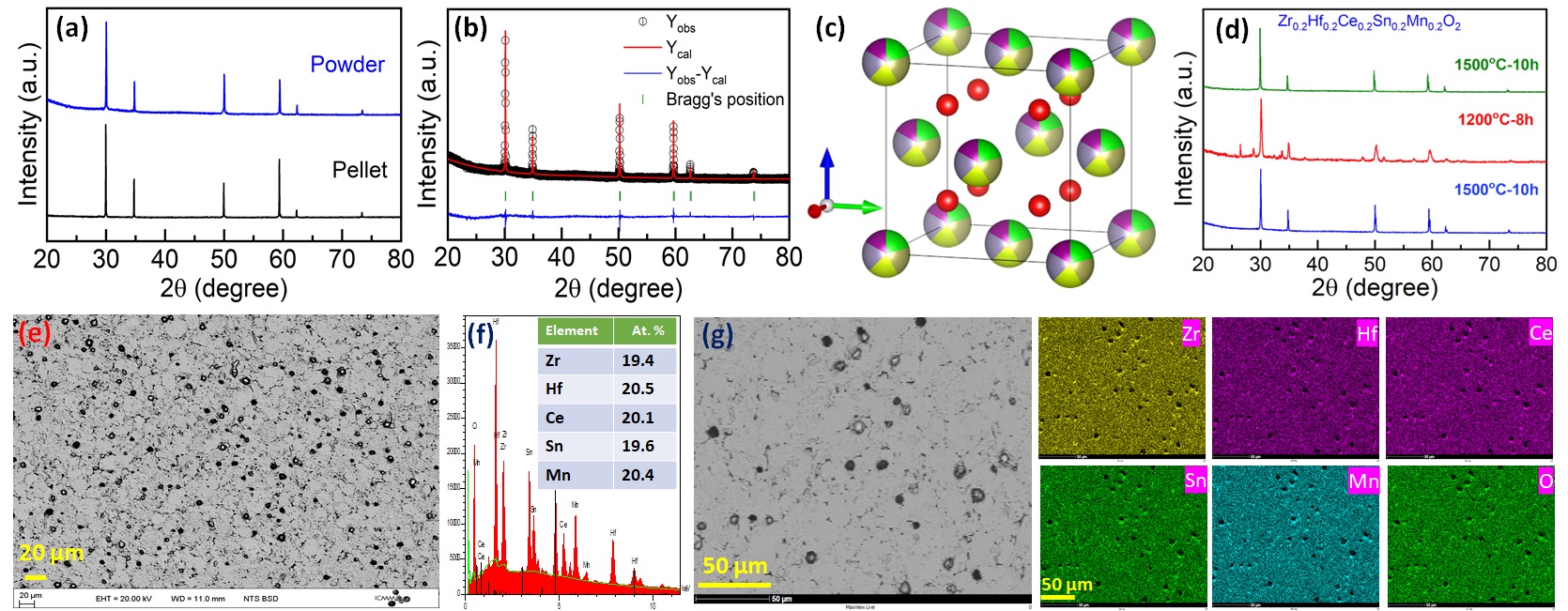}
  \caption{(a) X-ray diffraction pattern for Zr$_{0.2}$Hf$_{0.2}$Ce$_{0.2}$Sn$_{0.2}$Mn$_{0.2}$O$_{2-\delta}$ obtained on the pellet and powder. (b) Rietveld refinement pattern for Zr$_{0.2}$Hf$_{0.2}$Ce$_{0.2}$Sn$_{0.2}$Mn$_{0.2}$O$_{2-\delta}$. (c) Crystal structure obtained from Vesta software using structural parameters obtained from the Rietveld refinement. (d) The cyclic heat treatment demonstrates the entropy-stabilized nature of the sample. (e) Backscattered electron (BSE) image, (f) Energy-dispersive X-ray spectroscopy (EDS) pattern and (g) corresponding elemental mapping for Zr$_{0.2}$Hf$_{0.2}$Ce$_{0.2}$Sn$_{0.2}$Mn$_{0.2}$O$_{2-\delta}$.}
  \label{fig5}
\end{figure*}
\begin{figure}
\centering
   \includegraphics[width=0.99\linewidth]{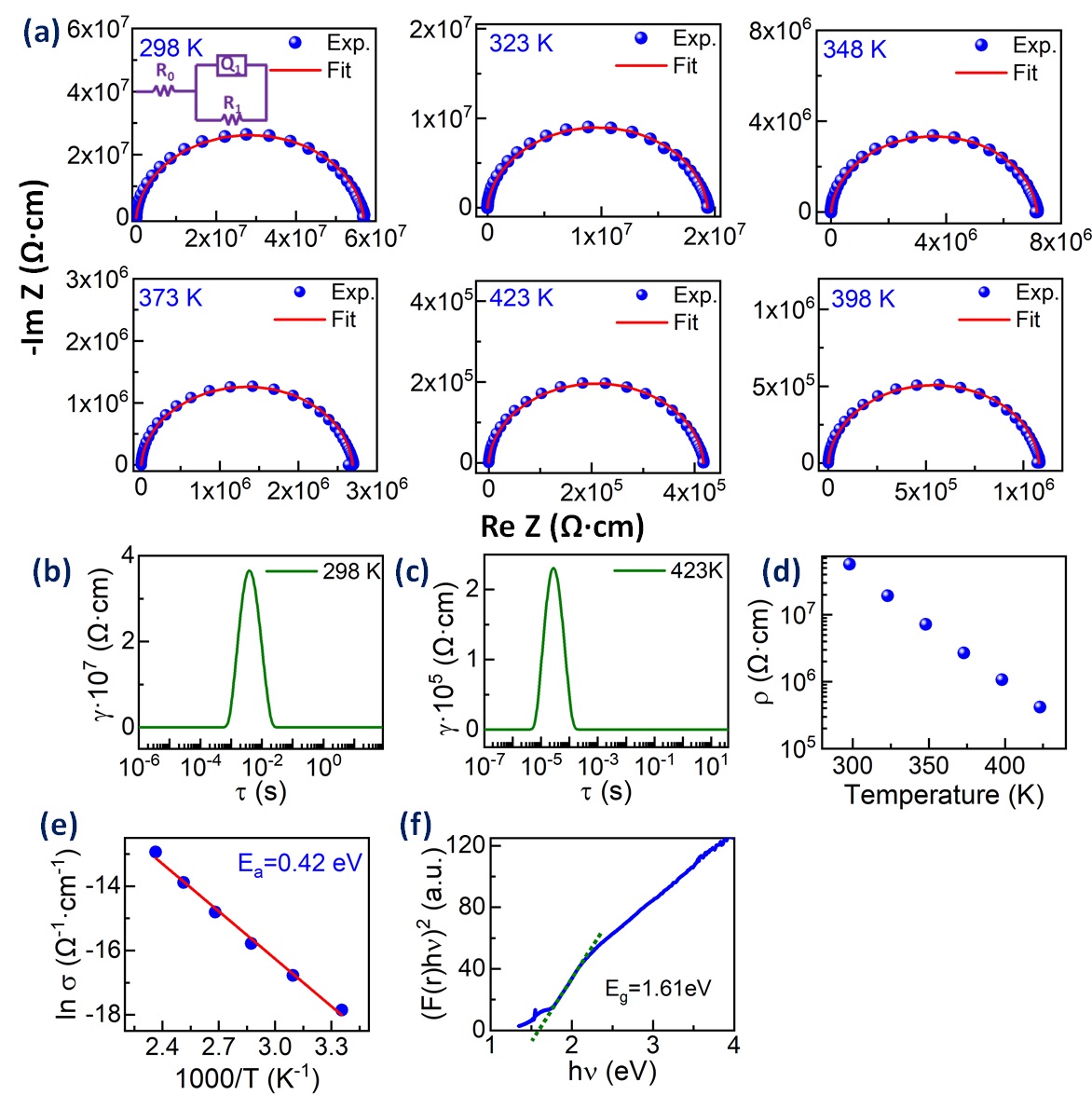}
  \caption{(a) Nyquist plot for Zr$_{0.2}$Hf$_{0.2}$Ce$_{0.2}$Sn$_{0.2}$Mn$_{0.2}$O$_{2-\delta}$ at different temperatures and corresponding single ZARC fit (the scheme is shown in inset). The distribution of relaxation time (DRT) plots for Zr$_{0.2}$Hf$_{0.2}$Ce$_{0.2}$Sn$_{0.2}$Mn$_{0.2}$O$_{2-\delta}$ (b) 298K (c) 423\,K, (d) The electrical resistivity ($\rho$) as a function of temperature, and (e) Arrhenius plot (ln$\sigma$ vs 1000/T) for Zr$_{0.2}$Hf$_{0.2}$Ce$_{0.2}$Sn$_{0.2}$Mn$_{0.2}$O$_{2-\delta}$, and (f) optical band gap obtained from UV-Vis spectra in diffuse reflectance mode.}
  \label{fig6}
\end{figure}
Further, SEM-EDS is carried out on the surface and the cross-section of the Zr$_{0.2}$Hf$_{0.2}$Ce$_{0.2}$Sn$_{0.2}$Ti$_{0.2}$O$_2$ pellet. The corresponding backscattered electron (BSE) image on the surface and bulk is shown in Fig.~\ref{fig3}(b-c). Interestingly, the surface shows no clear contrast whereas there is a contrast in the cross-section of the pellet depicting the multiphase nature of the bulk sample. The mapping of each element and corresponding EDS spectra are shown on the surface (Fig.~\ref{fig3}(d-e)) and bulk (Fig.~\ref{fig3}(f-g)) respectively. The homogenous distribution of elements may be seen on the surface, however, the cross-section mapping shows an inhomogeneous distribution, supporting the observation of multiphase by XRD. 
In order to further understand the different structures of the pellet and powder in the Sn-containing sample Zr$_{0.2}$Hf$_{0.2}$Ce$_{0.2}$Sn$_{0.2}$Ti$_{0.2}$O$_2$, X-ray photoelectron spectroscopy is carried out on the surface of the pellet and powder. Fig.~\ref{fig4} shows the core-level spectra for each element present in Zr$_{0.2}$Hf$_{0.2}$Ce$_{0.2}$Sn$_{0.2}$Ti$_{0.2}$O$_2$. It is seen that Zr, Hf, Ti, and Sn show similar core-level spectra for the pellet and powder and are in the 4+ charge state. However, it is interestingly observed that core-level spectra of Ce show two distinct charge states (Ce$^{4+}$ and Ce$^{3+}$) and are similar in both powder and pellet. The corresponding fitting is shown in Fig.~S8 and fitting details are listed in Table S3. The ratio Ce$^{3+}$/Ce$^{4+}$ was evaluated as proposed in refs\cite{pfau1994electronic, paparazzo2011curve} and found to be 1.4 for both locations. The fitting procedure, implying a Shirley-type background subtraction and mixed Gaussian-Lorentzian line shapes was performed using both the Avantage and CasaXPS software and shows similar results. However, this value is not precise as CeO$_2$ is known to reduce during X-ray exposure.\cite{qiu2006comparative, romeo1993xps} In general, fluorite structure is stabilized by oxygen deficiency, and to accommodate the overall charge neutrality Ce changes from Ce$^{4+}$ to Ce$^{3+}$, since all other elements are quite stable in 4+ charge states. Although the core-level spectra for all the cations are similar in bulk and surface, only core-level spectra for O-1s, shown in Fig.~\ref{fig4}, are observed to be different on the pellet and powder and may be the reason for such structural differences on the pellet and powder.\\  
There are several high-entropy fluorite oxides synthesized in the literature that possesses oxygen vacancies.\cite{djenadic2017multicomponent} Keeping these in mind, we then synthesized the Zr$_{0.2}$Hf$_{0.2}$Ce$_{0.2}$Sn$_{0.2}$Mn$_{0.2}$O$_{2-\delta}$ sample following the same synthesis route. The X-ray diffraction patterns, shown in Fig.~\ref{fig5}(a), depict the single-phase fluorite structure on the surface as well as in the bulk. The Rietveld refinement was performed of the bulk XRD pattern using Fullprof software\cite{rietveld1969profile} with \textit{Fm-3m} space group, and the refinement is shown in Fig.~\ref{fig5}(b). The lattice parameter obtained from the refinement for Zr$_{0.2}$Hf$_{0.2}$Ce$_{0.2}$Sn$_{0.2}$Mn$_{0.2}$O$_{2-\delta}$ is 5.136\AA, and the corresponding crystal structure drawn using Vesta software\cite{momma2011vesta} is shown in Fig.~\ref{fig5}(c). Fig.~\ref{fig5}(d) shows the X-ray diffraction pattern of the cyclic heat treatment performed on Zr$_{0.2}$Hf$_{0.2}$Ce$_{0.2}$Sn$_{0.2}$Mn$_{0.2}$O$_{2-\delta}$. It is seen that the sample is multiphase below a critical temperature and is reversed back to the single phase when sintered at 1500$^{\circ}$C followed by quenching, establishing the entropy-dominated phase stability of this sample. Therefore, Zr$_{0.2}$Hf$_{0.2}$Ce$_{0.2}$Sn$_{0.2}$Mn$_{0.2}$O$_{2-\delta}$ seems to be the first single-phase entropy-stabilized fluorite oxide. \\
Further, the backscattered electron image of Zr$_{0.2}$Hf$_{0.2}$Ce$_{0.2}$Sn$_{0.2}$Mn$_{0.2}$O$_{2-\delta}$ pellet (cross-section) is shown in Fig.~\ref{fig5}(e), and the corresponding energy-dispersive x-ray spectrum (EDS) in Fig.~\ref{fig5}(f). For the investigation of the microstructure, the surface of the pellet is polished using an automatic grinding/polishing machine. Due to polishing, the grains partly come out of the surface, as can be seen in the higher magnification image (Fig.~\ref{fig5}(g)). The atomic percentage of elements obtained from EDS is in line with the stoichiometric amount of precursors used for the synthesis. The elemental mapping for each element present in Zr$_{0.2}$Hf$_{0.2}$Ce$_{0.2}$Sn$_{0.2}$Mn$_{0.2}$O$_{2-\delta}$ is shown in Fig.~\ref{fig5}(g). The mapping is also performed on the cross-section of the pellet. The homogenous distribution of each element present in this entropy-stabilized Zr$_{0.2}$Hf$_{0.2}$Ce$_{0.2}$Sn$_{0.2}$Mn$_{0.2}$O$_{2-\delta}$ further confirms the single-phase nature of the sample. A comparison of Zr$_{0.2}$Hf$_{0.2}$Ti$_{0.2}$Sn$_{0.2}$Mn$_{0.2}$O$_{2-\delta}$ (Fig.~\ref{Fig2}(a)) and Zr$_{0.2}$Hf$_{0.2}$Ce$_{0.2}$Sn$_{0.2}$Mn$_{0.2}$O$_{2-\delta}$ (Fig.~\ref{fig5}(a)) shows that Ce seems to be essential to form single-phase fluorite structure, as also suggested by Chen et al. in a recent study.\cite{chen2023formation}\\
The charge transport properties in entropy-stabilized Zr$_{0.2}$Hf$_{0.2}$Ce$_{0.2}$Sn$_{0.2}$Mn$_{0.2}$O$_{2-\delta}$ sample are evaluated using Impedance spectroscopy (IS) techniques. The measurements were performed using ionically-blocking electrodes (Pt in the present case) on both sides of the pellet, as shown in Fig.~S12. The Bode plot consisting of the magnitude of the impedance |Z| and phase angle ($\phi$) as a function of frequency at different temperatures are shown in Fig.~S13(a-b). The Nyquist plots (-ImZ vs ReZ) for Zr$_{0.2}$Hf$_{0.2}$Ce$_{0.2}$Sn$_{0.2}$Mn$_{0.2}$O$_{2-\delta}$ at different temperatures are shown in Fig.~\ref{fig6}(a). Nyquist plots provide visual cues that help us analyze experimental data and understand the physical processes of the samples studied using IS measurement. Experimentally a perfect semicircle in the Nyquist plot is rarely seen due to the overlap of processes with similar time constants and hence an equivalent circuit model is widely used. However, it is seen that the analysis of IS using the corresponding equivalent circuits is very difficult due to the possibilities of overlap of different transport contributions. Therefore, the distribution of relaxation time (DRT) method has been promising to distinguish the characteristic relaxation times corresponding to different transport processes.\cite{liu2020deep, ciucci2019modeling} This method uses inverse Fourier transform to change the frequency-dependent IS data to a distribution of time constants to determine the relaxation times contributing to the impedance spectrum.\cite{schmidt2013distribution, steinhauer2017investigation}\\
The DRT plot obtained for Zr$_{0.2}$Hf$_{0.2}$Ce$_{0.2}$Sn$_{0.2}$Mn$_{0.2}$O$_{2-\delta}$ to identify the number of processes involved in the transport process is shown in Fig.~\ref{fig6}(b,c) at 298K and 423K for Zr$_{0.2}$Hf$_{0.2}$Ce$_{0.2}$Sn$_{0.2}$Mn$_{0.2}$O$_{2-\delta}$. DRT plot gives one dominant contribution and hence all the Nyquist plots are fitted with a single ZARC model, where a ZARC is defined as a parallel combination of a resistor (R) and constant phase element (Q) as shown in the inset of Fig.~\ref{fig6}a. As seen from the Nyquist plot, this sample does not show a capacitive tail at lower frequencies with Pt-blocking electrodes. Thus, it is highly possible that the ionic conduction is negligible in this sample and that the only contribution to the electrical transport is electronic. The electrical resistivities obtained from the fitting are shown in Fig.~\ref{fig6}(d), which is in line with the two-probe electrical resistivity measurement performed on this sample (not shown). The fitting parameters obtained from the ZARC fit are shown in Table S5. Further, these electrical resistivities are used to estimate the activation energy from the Arrhenius plot, as shown in Fig.~\ref{fig6}(e). The activation energy (Ea) is 0.42 eV which is quite lower than the activation energy reported for Zr$_{0.2}$Hf$_{0.2}$Ce$_{0.2}$Sn$_{0.2}$Ti$_{0.2}$O$_2$.\cite{chen2018five}\\
The UV--Visible spectroscopy measurement is carried out at 300\,K in diffuse reflectance spectra (DRS) mode to find the optical band gap for Zr$_{0.2}$Hf$_{0.2}$Ce$_{0.2}$Sn$_{0.2}$Mn$_{0.2}$O$_{2-\delta}$. The Tauc method for estimating band gap, based on the assumption of energy-dependent absorption coefficient ($\alpha$), is expressed as $(\alpha h\nu)^{1/n}$=$A(h\nu-E_g)$, where $h$ is Planck constant, $\nu$ is the photon's frequency, E$_g$ is the optical band gap, and $A$ is a constant. The factor $n$ depends on the nature of electronic transition and is taken 1/2 and 2 for direct and indirect transition band gaps, respectively. For the determination of band gap in DRS mode, Kubelka, and Munk suggested the transformation of measured reflectance spectra to the corresponding absorption spectra by applying the Kubelka-Munk function (F(r)), defined as $F(r)=\frac{(1-R)^2}{2R}$ where $R$ is the reflectance of an infinitely thick specimen.\cite{makula2018correctly} Hence, the above equation reduces to  $(F(r)·h \nu)^{1/n}$=$A(h\nu-E_g)$. The direct band gap for Zr$_{0.2}$Hf$_{0.2}$Ce$_{0.2}$Sn$_{0.2}$Mn$_{0.2}$O$_{2-\delta}$ is calculated using $(F(r)h\nu)^2$  vs. h$\nu$ curve and is shown in Fig.~\ref{fig6}(f). It is noted that the optical band gap for Zr$_{0.2}$Hf$_{0.2}$Ce$_{0.2}$Sn$_{0.2}$Ti$_{0.2}$O$_2$ is 3.05 eV (Fig.~S11) and it reduces to 1.61 eV for Zr$_{0.2}$Hf$_{0.2}$Ce$_{0.2}$Sn$_{0.2}$Mn$_{0.2}$O$_{2-\delta}$.\\ 
Next, we explore more compositions by following the selection rules which include the ionic radii and corresponding oxidation state of cations criteria following Pauling rules (for structure and local coordination of cations). These criteria, similar to conventional oxide compounds, are decisive for establishing the phase stability of high-entropy oxides, although exceptions exist.\cite{jiang2018new, edalati2020photocatalytic} Accordingly, the average value of cationic radii $\left(\Bar{r}=\frac{1}{N}\left(\sum_{i=1}^{N}r_i\right)\right)$, standard deviation $\left(s=\sqrt{\frac{\sum_{i=1}^{N}(r_i-\Bar{r})^2}{(N-1)}}\right)$ with $N$ is the number of cations, has been used to synthesize high-entropy oxides, and are presented in Table~\ref{table1}. It is noted that s $>$ 0.095 does not result in single-phase fluorite oxide, as predicted by Spiridigliozzi et al,\cite{spiridigliozzi2021simple}, as seen in Zr$_{0.2}$Hf$_{0.2}$Ce$_{0.2}$Sn$_{0.2}$Ti$_{0.2}$O$_2$. The nature of the diffraction pattern obtained on the surface(S) and bulk (B) for the samples studied here is also depicted in Table~\ref{table1}. Several optimizations in these selection rules result in three new compositions that lead to the single-phase fluorite structure and are entropy stabilized. The X-ray diffraction patterns for these three new compositions are shown in Fig.~\ref{fig7}(a). These samples exhibit the same XRD pattern in both pellet and powder form. Zr$_{0.2}$Hf$_{0.2}$Ti$_{0.2}$Mn$_{0.2}$Ce$_{0.2}$O$_{2-\delta}$ shows a dominant fluorite structure with a minor CeO$_2$ phase (marked with $\ast$ in Fig.~\ref{fig7}(a)) in very little amount. This minor CeO$_2$ phase present in the first sample can be suppressed by reducing the cerium content in the sample and synthesizing Zr$_{0.225}$Hf$_{0.225}$Ti$_{0.225}$Mn$_{0.225}$Ce$_{0.1}$O$_{2-\delta}$, which shows a single-phase fluorite structure. Further, Zr$_{0.2}$Hf$_{0.2}$Ti$_{0.2}$Mn$_{0.2}$Ce$_{0.1}$Ta$_{0.05}$Fe$_{0.05}$O$_{2-\delta}$ also shows a single-phase fluorite structure with an \textit{Fm-3m} space group, showing that a 4+ element can be partially substituted by an equimolar combination of a 3+ and a 5+ element. The temperature of the sintering and the quenching procedure has been optimized to achieve single-phase fluorite oxide. The synthesis of these samples does not require intermediate calcination steps. Zr$_{0.2}$Hf$_{0.2}$Ti$_{0.2}$Mn$_{0.2}$Ce$_{0.2}$O$_{2-\delta}$ sample sintered at 1400$^{\circ}$C followed by furnace cooling shows a mixture of fluorite and pyrochlore phases, as shown in Fig.~S3, suggesting that the sample has to be quenched faster for single phase formation. Further, following the same sintering condition, the sample was quenched in the air (crucibles removed from the furnace and kept outside) and the diffraction pattern (shown in Fig.~S4) exhibits  a minor shoulder for each Bragg’s peak, which disappears when the quenching is done on a metallic plate. The difference between the quenching done in the air and on the metallic plate is shown in Fig.~S5. A symmetric XRD peak was obtained for the sample quenched on the metallic plate. This suggests that quenching the sample in the air is not sufficient to get single-phase fluorite oxide. The entropy-stabilization is however established in these samples having minor impurity phases as well (Fig.~S5). Also, the temperature of sintering is further optimized to 1380$^{\circ}$C to achieve a uniform pellet.\\
\begin{figure*}
\centering
   \includegraphics[width=0.99\linewidth]{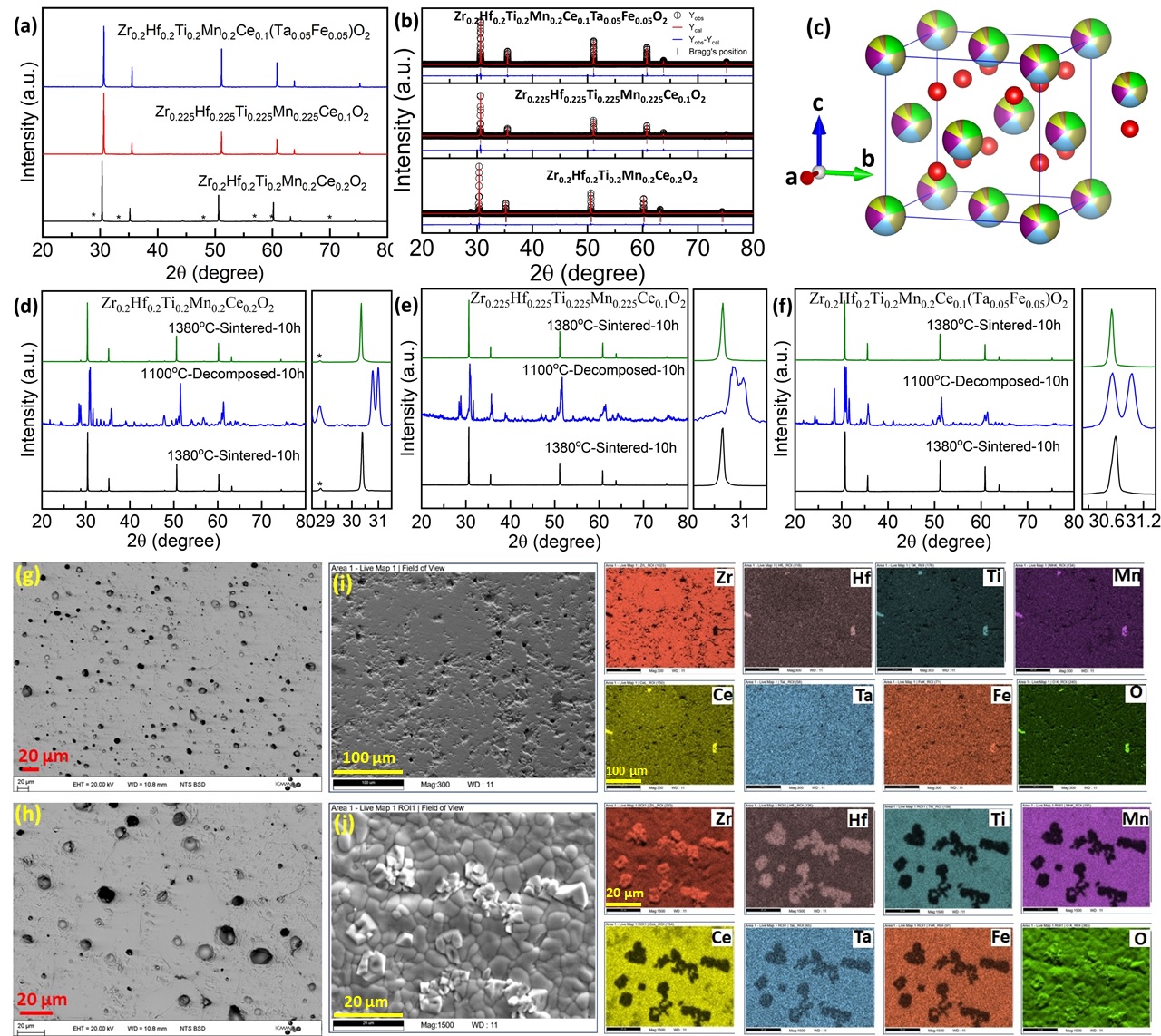}
  \caption{(a) X-ray diffraction pattern, (b) Rietveld refinement pattern for three single-phase fluorite oxides. The minor CeO2 peaks present in Zr$_{0.2}$Hf$_{0.2}$Ti$_{0.2}$Mn$_{0.2}$Ce$_{0.2}$O$_{2-\delta}$, are marked with the star-shaped symbol (*) in (a). (c) The crystal structure for Zr$_{0.2}$Hf$_{0.2}$Ti$_{0.2}$Mn$_{0.2}$Ce$_{0.1}$Ta$_{0.05}$Fe$_{0.05}$O$_{2-\delta}$ is drawn using the refinement parameters employing Vesta software. (d-f) The cyclic heat treatment demonstrating entropy-stabilization of the fluorite is shown for all three samples. (g-h) Backscattered electron (BSE) image Zr$_{0.2}$Hf$_{0.2}$Ti$_{0.2}$Mn$_{0.2}$Ce$_{0.1}$Ta$_{0.05}$Fe$_{0.05}$O$_{2-\delta}$, Elemental mappings for Zr$_{0.2}$Hf$_{0.2}$Ti$_{0.2}$Mn$_{0.2}$Ce$_{0.1}$Ta$_{0.05}$Fe$_{0.05}$O$_{2-\delta}$, (i) single-phase: sintered at 1380$^{\circ}$C followed by quenching (j) decomposed sample: sintered at 1300$^{\circ}$C followed by slow cooling.}
  \label{fig7}
\end{figure*}
The XRD pattern of these three samples is further analyzed using Rietveld refinement employing Fullprof software considering the \textit{Fm-3m} space group. The refinement pattern for each phase is shown in Fig.~\ref{fig7}(b). The lattice parameter for Zr$_{0.2}$Hf$_{0.2}$Ti$_{0.2}$Mn$_{0.2}$Ce$_{0.2}$O$_{2-\delta}$  is 5.095 \AA, Zr$_{0.225}$Hf$_{0.225}$Ti$_{0.225}$Mn$_{0.225}$Ce$_{0.1}$O$_{2-\delta}$ is 5.052 \AA, and Zr$_{0.2}$Hf$_{0.2}$Ti$_{0.2}$Mn$_{0.2}$Ce$_{0.1}$Ta$_{0.05}$Fe$_{0.05}$O$_{2-\delta}$  is 5.051 \AA and are consistent with the change in the ionic radii of the cations for each composition.\cite{shannon1976revised} The crystal structure for Zr$_{0.2}$Hf$_{0.2}$Ti$_{0.2}$Mn$_{0.2}$Ce$_{0.1}$Ta$_{0.05}$Fe$_{0.05}$O$_{2-\delta}$ was drawn from Vesta software using the structural parameters (lattice parameter, atomic position, and occupancy) obtained from the refinement and is shown in Fig.~\ref{fig7}(c). The entropy stabilization in these systems has been investigated through the cyclic heat treatment as shown in Fig.~\ref{fig7}(d-f) for each composition. The reversible observation of the single-phase $\rightarrow$ multiphase $\rightarrow$ single-phase diffraction pattern of these samples with changing the sintering temperature below and above a critical temperature confirms that the thermodynamic stability of these samples is dominated by configurational entropy. The cyclic heat treatment at several temperatures in these systems is also shown in Fig.~S6, showing that this critical temperature is between 1300$^{\circ}$C and 1380$^{\circ}$C. As the use of a sintering temperature larger than 1380$^{\circ}$C leads to a degradation of the pellets (due to Mn melting), it shows that the temperature window that can be used to obtain these new compositions is rather narrow.\\
Surface morphology and chemical homogeneity studies of the entropy-stabilized samples are carried out using the SEM-EDS technique. The microstructure shows a dense morphology of the samples sintered at 1380$^{\circ}$C having a relative density greater than 93\%. Fig.~\ref{fig7}(g,h) show the surface morphology for Zr$_{0.2}$Hf$_{0.2}$Ti$_{0.2}$Mn$_{0.2}$Ce$_{0.1}$Ta$_{0.05}$Fe$_{0.05}$O$_{2-\delta}$ in backscattered electron (BSE) mode. As mentioned earlier, the polishing of the pellets results in the removal of grains from the surface which can be in a higher magnification image (Fig.~\ref{fig7}(h)). The sample without polishing looks quite dense, as shown in Fig.~\ref{fig7}(j). The uniform surface morphology further supports the single-phase nature of the sample. EDS mapping shows the compositional homogeneity as all the constituent elements (Zr, Hf, Ti, Mn, Ce, Ta, Fe, and O) were found to be uniformly distributed across the investigated area of the microstructure, as shown in Fig.~\ref{fig7}(i) for single-phase Zr$_{0.2}$Hf$_{0.2}$Ti$_{0.2}$Mn$_{0.2}$Ce$_{0.1}$Ta$_{0.05}$Fe$_{0.05}$O$_{2-\delta}$. Figure 7(j) shows the surface morphology (BSE image) and corresponding elemental map for the Zr$_{0.2}$Hf$_{0.2}$Ti$_{0.2}$Mn$_{0.2}$Ce$_{0.1}$Ta$_{0.05}$Fe$_{0.05}$O$_{2-\delta}$,  sample decomposed at 1300$^{\circ}$C for 10 hours. The multiple phases and non-uniform distribution observed in the elemental map further confirm the decomposed nature of the sample.
\begin{table*}
\small
  \caption{\ Different high-entropy binary oxide synthesized in the present study with corresponding sintering temperature (T) is shown. The average ionic radii ($\Bar{r}$), standard deviation ($s$), sintering temperature ($T$), configurational entropy, and the nature of the XRD pattern obtained from the surface (S) and powder (B) are shown for comparison (F: Fluorite (\textit{Fm-3m}), O: Orthorhombic (\textit{Pbcn}), P: Pyrochlore (\textit{Fm-3m}). The elements in the parenthesis are in the equimolar ratio. XRD pattern for (ZrHfTiCeY)O$_{2-\delta}$, (ZrHfCeY)O$_{2-\delta}$, (ZrHfCeYGd)O$_{2-\delta}$, (ZrHfCeYYb)O$_{2-\delta}$, (ZrHfCeYbGd)O$_{2-\delta}$ are shown in the Fig.~S7 of the supporting information.}
  \label{table1}
  \begin{tabular*}{0.9\textwidth}{@{\extracolsep{\fill}}lllllllll}
    \hline
   Sample & $\Bar{r}$ & $s$ & T & S$_{conf}$ & Entropy & XRD & XRD & Ref.\\
    & \AA &  & ($^{\circ}$C) &  & stabilized & Surface & Bulk & \\
    \hline
     (ZrHfTiSnCe)O$_2$ & 0.761	& 0.135 &	1500 &	1.61R	& Yes : Surface	& F	& F+O & \cite{chen2018five, he2021four, ding2022hf0}\\
     (ZrHfTiSnMn)O$_{2-\delta}$ &	0.733 &	0.087	& 1500 &	1.61R	& Yes : Surface	& F &	F+O	& This work\\
     (ZrHfTiSn)0.8(CeY)0.2O$_{2-\delta}$ &	0.754	& 0.138 &	1500	& 1.75R	& Yes : Surface	& F &	F+O	& This work\\
     (ZrHfTiSnY)O$_{2-\delta}$ &	0.747	& 0.109	& 1500	& 1.61R	& Yes : Surface	& F	& F+O	& This work\\
     (ZrHfCeSnMn)O$_{2-\delta}$	& 0.806	& 0.105	& 1500	& 1.61R	& Yes	& F	& F	& This work\\
     (ZrHfTiCeY)O$_{2-\delta}$	& 0.803	& 0.140	& 1500	& 1.61R &	No	& F+P	& F+P	& \cite{gild2018high}\\
     (ZrHfCeY)O$_{2-\delta}$	& 0.853	& 0.099 &	1500 &	1.29R	& No	& F	& F	& \cite{gild2018high}\\

(ZrHfCeYGd)O$_{2-\delta}$	& 0.87	& 0.094	& 1500	& 1.61R	& No	& F+P	& F+P	& \cite{gild2018high}\\

(ZrHfCeYYb)O$_{2-\delta}$	& 0.855 &	0.086	& 1500& 1.61R	&No	&F+P	&F+P	&\cite{gild2018high}\\

(ZrHfCeYbGd)O$_{2-\delta}$	&0.863	&0.093	&1500	&1.61R	&No	&F+P	&F+P	&\cite{gild2018high}\\
(ZrHfTiNbY)O$_{2-\delta}$	&0.737	&0.118	&1500&	1.61R	&No	&M&	M	&This work\\
(ZrHfTiMnCe)O$_{2-\delta}$	&0.789	&0.132	&1380	&1.61R	&Yes&	F	&F	&This work\\
((ZrHfTiMn)$_{0.9}$Ce$_{0.1}$)O$_{2-\delta}$&	0.766&	0.134	&1380	&1.57R	&Yes&	F&	F&	This work\\
(ZrHfTiMn)$_{0.8}$Ce$_{0.1}$(TaFe)$_{0.1}$O$_{2-\delta}$&	0.756	&0.128&	1380	&1.66R	&Yes	&F	&F	&This work\\
    \hline
  \end{tabular*}
\end{table*}
X-ray photoelectron spectroscopy (XPS) studies are carried out in order to determine the charge states of each element in these entropy-stabilized systems. Full range spectra of the samples are shown in Fig.~S9. The core level spectra, displayed in Fig.~\ref{fig8}(a-f), reveals the presence of Zr(4+), Hf(4+), Ti(4+), Mn(2+), Ce(4+), Ta(5+), Fe(3+) (corresponding binding energies shows in Table~S2) and are similar in all three samples. Interestingly Mn shows a 2+ valence state (see Fig.~S10 and Table~S4 of the SI), which is confirmed by the satellite peak characteristic to MnO (646.6 eV) in Fig.~\ref{fig8}(e) and possibly accommodates oxygen vacancies present in the sample as shown in Fig.~\ref{fig8}(g).\cite{junta1994manganese} This reduction of Mn from Mn$^{4+}$ to Mn$^{2+}$ increases the cationic radius of Mn from 0.53 \AA (Mn$^{4+}$) to 0.67$^{LS}$ \AA, 0.83$^{HS}$ \AA (Mn$^{2+}$). This increase in the ionic radii is close to the average ionic radii of other elements. Fig.~\ref{fig8}(h) shows a comparison of core-level spectra for Ce in Zr$_{0.2}$Hf$_{0.2}$Ti$_{0.2}$Mn$_{0.2}$Ce$_{0.2}$O$_{2-\delta}$ and Zr$_{0.2}$Hf$_{0.2}$Ti$_{0.2}$Sn$_{0.2}$Ce$_{0.2}$O$_2$. It is seen that Ce shows two distinct charge states in the Sn-containing sample (Table S3), however, it retains its Ce$^{4+}$ charge state in Mn-containing samples. This is possibly the reason for the different structural behavior on the surface and bulk of Sn-containing samples. Therefore, this also indicates that Mn is an essential element to synthesize high-entropy fluorite oxides, as the reduction of Mn into a 2+ state enables Ce to remain in its 4+ charge state.\\

\begin{figure*}
\centering
   \includegraphics[width=0.99\linewidth]{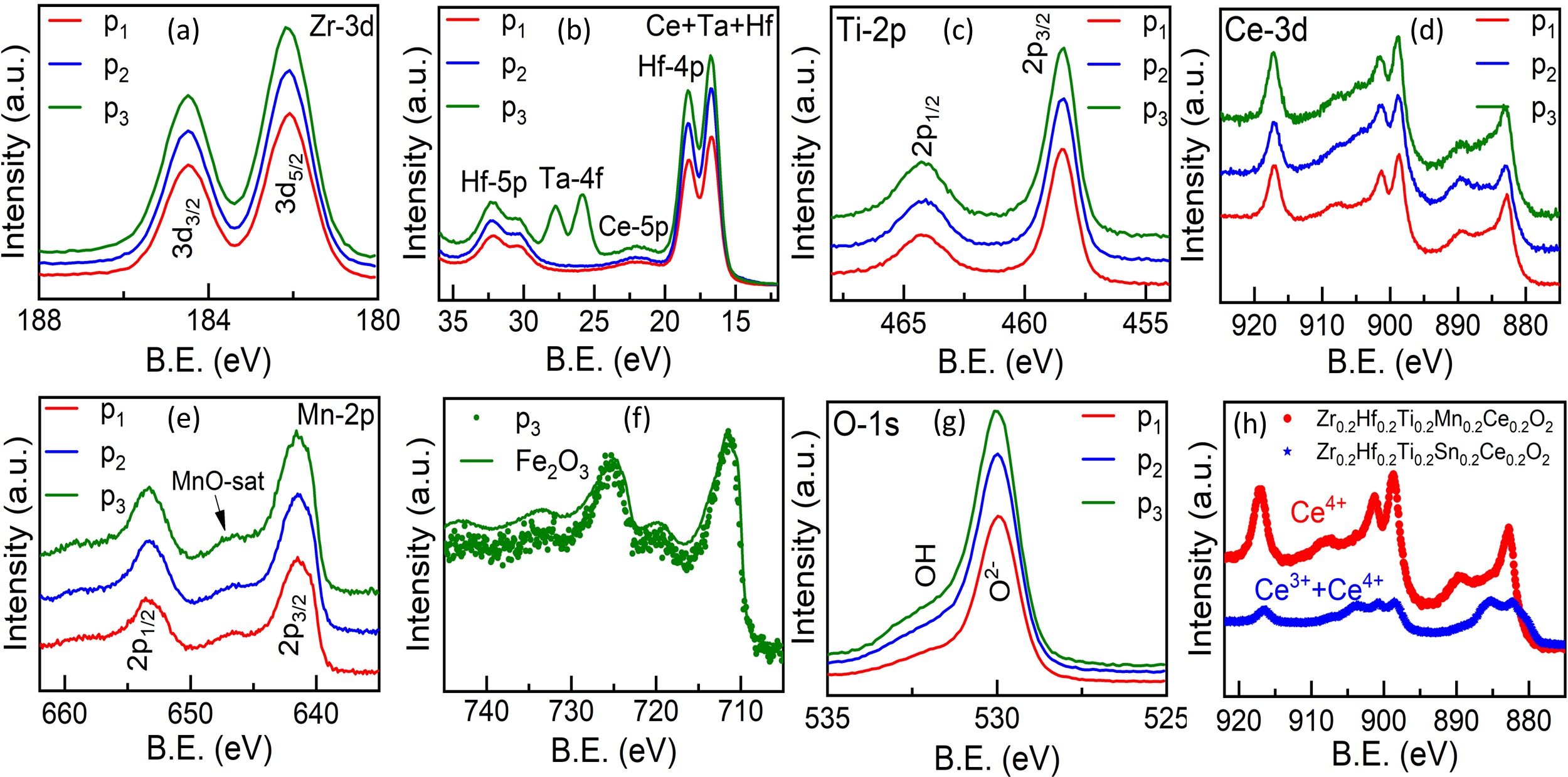}
  \caption{(a-g) Core-level X-ray photoelectron spectroscopy (XPS) spectra for the elements present in entropy-stabilized fluorite oxides. p$_1$: Zr$_{0.2}$Hf$_{0.2}$Ti$_{0.2}$Mn$_{0.2}$Ce$_{0.2}$O$_{2-\delta}$, p$_2$: Zr$_{0.225}$Hf$_{0.225}$Ti$_{0.225}$Mn$_{0.225}$Ce$_{0.1}$O$_{2-\delta}$ and p$_3$: Zr$_{0.2}$Hf$_{0.2}$Ti$_{0.2}$Mn$_{0.2}$Ce$_{0.1}$Ta$_{0.05}$Fe$_{0.05}$O$_{2-\delta}$. (h) comparison of Ce core-level XPS spectra for Zr$_{0.2}$Hf$_{0.2}$Ti$_{0.2}$Mn$_{0.2}$Ce$_{0.2}$O$_{2-\delta}$ and Zr$_{0.2}$Hf$_{0.2}$Ti$_{0.2}$Sn$_{0.2}$Ce$_{0.2}$O$_{2-\delta}$.}
  \label{fig8}
\end{figure*}
The Nyquist plots for all the samples are shown in Fig.~S14 and the corresponding Bode plot consisting of the magnitude of the impedance |Z| and phase angle ($\phi$) against frequency at different temperatures are shown in Fig.~S15. The measurements were performed in the same experimental conditions as mentioned above. The distribution of relaxation time (DRT) method has been used to determine the relaxation times contributing to the impedance spectrum.\cite{steinhauer2017investigation, schmidt2013distribution} The DRT plots are obtained for all the samples to identify the number of processes involved in the transport process and are shown in Fig.~S14(b,d,f) at different temperatures. It is noted that Zr$_{0.2}$Hf$_{0.2}$Ti$_{0.2}$Mn$_{0.2}$Ce$_{0.2}$O$_{2-\delta}$, shows two clearly distinguishable processes across the temperature range studies and hence two ZARC model was used to fit the corresponding IS data. The two contributions might result from electronic contribution due to Zr$_{0.2}$Hf$_{0.2}$Ti$_{0.2}$Mn$_{0.2}$Ce$_{0.2}$O$_{2-\delta}$ and minor CeO$_2$ phase present in the sample. The second contribution may also originate from grain boundaries. Zr$_{0.225}$Hf$_{0.225}$Ti$_{0.225}$Mn$_{0.225}$Ce$_{0.1}$O$_{2-\delta}$ shows only one process and hence single ZARC is used to estimate the transport parameters that are dominated by electronic conduction. Similar single-ZARC model is used to fit the IS data for Zr$_{0.2}$Hf$_{0.2}$Ti$_{0.2}$Mn$_{0.2}$Ce$_{0.1}$Ta$_{0.05}$Fe$_{0.05}$O$_{2-\delta}$. In general, the DRT analysis provides the relaxation time and corresponding resistances (ionic, grain boundary or electronic) of the processes involved, however, these resistances are not the actual resistance but they represent the linear combination of these. Hence these values are used as a starting point to fit the IS data using ZARC fit. The Nyquist plots are fitted using fit3Zarcs software\cite{dragoezarc3fit} using the appropriate equivalent circuit as shown in the respective insets of Fig.~S14. Fig.~S14(a) shows the Nyquist plot for Zr$_{0.2}$Hf$_{0.2}$Ti$_{0.2}$Mn$_{0.2}$Ce$_{0.2}$O$_{2-\delta}$ at four different temperatures, which are fitted using two ZARC elements (each consisting of a resistor (R) and constant phase element (Q) in parallel as shown in Fig.~S14(a). It is observed that the impedance changes by one order of magnitude for a change in temperature from 298K to 373K. The resistivity values obtained from the fitting at 298K are R1=1.81E8 $\Omega$·cm \& R2=5.18E7 $\Omega$·cm and decrease to 7.41E6 $\Omega$·cm \& 2.75E6 $\Omega$·cm respectively at 373K, indicating thermally activated behavior with an activation energy of 0.34 eV and 0.42 eV respectively (Fig.~S16 (a-b)). Zr$_{0.225}$Hf$_{0.225}$Ti$_{0.225}$Mn$_{0.225}$Ce$_{0.1}$O$_{2-\delta}$ was fitted using the proposed model as shown in Fig.~S14(b). The refined value of electrical resistivity at 298K is 3.08E8 $\Omega$·cm and it reduces to 1.44E7 $\Omega$·cm at 373K. The activation energy obtained from the linear fit of the Arrhenius plot is 0.42 eV (Fig.~S16(c)). The fitting of Nyquist plot for Zr$_{0.2}$Hf$_{0.2}$Ti$_{0.2}$Mn$_{0.2}$Ce$_{0.1}$Ta$_{0.05}$Fe$_{0.05}$O$_{2-\delta}$ shows a change in the resistance from 3.37E8 $\Omega$·cm at 298K to 3.50E6 $\Omega$·cm at 423K with an activation energy of 0.4 eV (Fig.~S16(d)). The detailed transport parameters obtained from the fitting for each sample are shown in Table S6-S8.\\
The magnetic measurements (M-T) are carried out in zero-field cooling (ZFC) and field cooling (FC) mode over a temperature range of 2 K-350 K. It is seen that all three samples show similar magnetic behavior with a paramagnetic contribution over a wide temperature range (Fig.~S17).   For all compositions, the Curie-Weiss fit of the 1/$\chi$ vs T curve evidences antiferromagnetic interactions with a Curie-Weiss temperature ($\theta_{cw}$) of -28K, -30K and -43K respectively for Zr$_{0.2}$Hf$_{0.2}$Ti$_{0.2}$Mn$_{0.2}$Ce$_{0.2}$O$_{2-\delta}$, Zr$_{0.225}$Hf$_{0.225}$Ti$_{0.225}$Mn$_{0.225}$Ce$_{0.1}$O$_{2-\delta}$ and Zr$_{0.2}$Hf$_{0.2}$Ti$_{0.2}$Mn$_{0.2}$Ce$_{0.1}$Ta$_{0.05}$Fe$_{0.05}$O$_{2-\delta}$. Unexpectedly, the paramagnetic moment extracted from the Curie-Weiss fit leads to unrealistic values, much larger than those expected from free Mn2+ in high-spin configuration (even taking into account a possible moderate contribution from Ce$^{3+}$ and Fe$^{3+}$), from 8.48 $\mu_B$ in Zr$_{0.2}$Hf$_{0.2}$Ti$_{0.2}$Mn$_{0.2}$Ce$_{0.2}$O$_{2-\delta}$ to 8.81 $\mu_B$ in Zr$_{0.2}$Hf$_{0.2}$Ti$_{0.2}$Mn$_{0.2}$Ce$_{0.1}$Ta$_{0.05}$Fe$_{0.05}$O$_{2-\delta}$ and 8.94 $\mu_B$ in Zr$_{0.225}$Hf$_{0.225}$Ti$_{0.225}$Mn$_{0.225}$Ce$_{0.1}$O$_{2-\delta}$. The M-H curve shown for these three samples at 2K (Fig.~S17) shows a rise in magnetization with the magnetic field, here again to large values without reaching saturation even after $\pm$5T. At this stage, we do not have any plausible explanation to explain these unexpected values, which are obtained for all three samples.\\
\begin{figure}
\centering
   \includegraphics[width=0.99\linewidth]{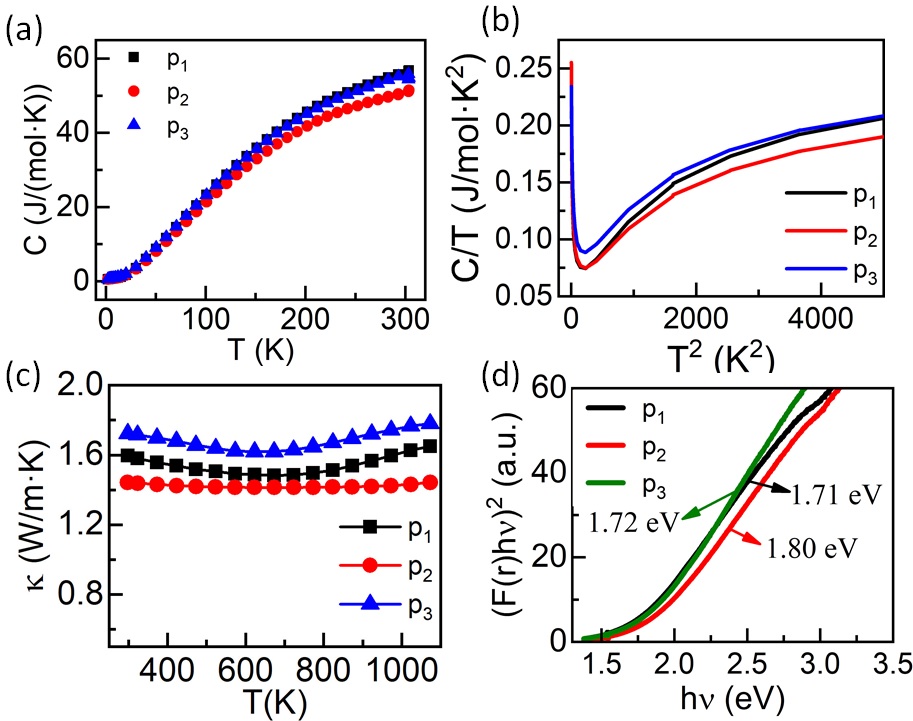}
  \caption{(a) specific heat capacity (C) as a function of temperature, (b) C/T vs T$^2$ (c) temperature-dependent thermal conductivity ($\kappa$), and (d) optical band-gap calculated using UV-Visible spectra for entropy-stabilized samples, where p$_1$: Zr$_{0.2}$Hf$_{0.2}$Ti$_{0.2}$Mn$_{0.2}$Ce$_{0.2}$O$_{2-\delta}$, p$_2$: Zr$_{0.225}$Hf$_{0.225}$Ti$_{0.225}$Mn$_{0.225}$Ce$_{0.1}$O$_{2-\delta}$ and p$_3$: Zr$_{0.2}$Hf$_{0.2}$Ti$_{0.2}$Mn$_{0.2}$Ce$_{0.1}$Ta$_{0.05}$Fe$_{0.05}$O$_{2-\delta}$.}
  \label{fig9}
\end{figure}
Specific heat capacity as a function of temperature is shown in Fig.~\ref{fig9}(a). As expected, the absolute value of the specific heat does not significantly differ from one composition to another. The room temperature value, close to 55 J·mol$^{-1}$·K$^{-1}$, is very similar to the value reported for ZrO$_2$\cite{degueldre2003specific} and constitutes a signature of a large Debye temperature.  The Debye temperature obtained for the present samples is close to 550 K and is consistent with the other reports for ZrO$_2$ based systems.\cite{degueldre2003specific}  Also, C$_p$ vs T data shows no anomaly in the whole temperature range. However, a plot of C$_p$/T vs. T$^2$ at low temperature evidences a strong upturn below $\sim$14\,K (Fig.~\ref{fig9}(b)), characteristic of a Schottky anomaly. This Schottky anomaly could originate either from the presence of magnetic ions in the compounds, although no magnetic ordering is expected even at very low temperatures due to the limited concentration of magnetic elements, or to a quadrupole interaction linked to the presence of 20\% to 22.5\% of Hf in a distorted environment in the samples.\cite{lawless1980low} 
The thermal conductivity ($\kappa$) studied over a wide range of temperatures (298\,K to 1073\,K) for these three entropy-stabilized fluorite oxides is shown in Fig.~\ref{fig9}(c). The temperature-dependent thermal diffusivity (D) is shown in Fig.~S18. At a first glance, $\kappa$ seems to decrease slightly with a rise in temperature up to 650\,K, above which it increases slowly. However, it should be noted that $\kappa$ has been calculated using the specific heat capacity from the Dulong-Petit law. Since the Debye temperature for these systems is close to $\sim$550K, there is an overestimation of C$_p$, and thus of $\kappa$ up to circa 600\,K. The thermal conductivity for all three samples is in the range of 1.42 W·m$^{-1}$·K$^{-1}$ to 1.73 W·m$^{-1}$·K$^{-1}$ at 298\,K and is significantly lower than the thermal conductivity values of binary oxides.\cite{suzuki2019thermal} This reduction as compared to the binary oxides can be explained by the random distribution of several cations with different masses and ionic radii that results in enhanced scattering of phonons, which leads to a reduced lattice contribution to the thermal conductivity. As these materials possess poor electronic thermal conductivity (due to small electrical conductivity) at room temperature, the reduction in phonon thermal conductivity reduces the total thermal conductivity. The slight upward rise in $\kappa$ at high temperatures may be due to the rise in electrical conductivity that enhances the electronic contribution to the thermal conductivity or due to photon thermal conductivity as reported recently\cite{zhang2023exploring} that may also enhance the total thermal conductivity at higher temperatures. UV-Visible spectroscopy in diffuse reflectance spectra (DRS) mode was performed at 300\,K to find the optical band gap in these entropy-stabilized samples, as discussed above. The optical band gap (assuming a direct band gap) obtained for our three new entropy-stabilized samples is shown in Fig.~\ref{fig9}(d). It is noted that the optical band gap for Zr$_{0.2}$Hf$_{0.2}$Ce$_{0.2}$Ti$_{0.2}$Sn$_{0.2}$O$_2$ is 3.05 eV (Fig.~S11) and it reduces to 1.72 eV for Zr$_{0.2}$Hf$_{0.2}$Ti$_{0.2}$Mn$_{0.2}$Ce$_{0.1}$Ta$_{0.05}$Fe$_{0.05}$O$_{2-\delta}$. This reduction in band gap from near UV to visible could probably be linked to the introduction of 3d$^5$ Mn$^{2+}$ ions in these new entropy-stabilized samples, which could make them promising for photocatalytic application due to the absorption of wavelength in the visible region.\\
\section{Conclusion}
This study presents a set of novel entropy-stabilized fluorite oxides using a combination of diffraction and spectroscopic techniques. First, we explore that entropy-stabilized fluorite oxide containing Sn as one of the elements viz. (Zr$_{0.2}$Hf$_{0.2}$Ti$_{0.2}$Sn$_{0.2}$Ce$_{0.2}$)O$_2$, (Zr$_{0.2}$Hf$_{0.2}$Ti$_{0.2}$Sn$_{0.2}$Mn$_{0.2}$)O$_{2-\delta}$, and (Zr$_{0.2}$Hf$_{0.2}$Ti$_{0.2}$Sn$_{0.2}$Ce$_{0.1}$Y$_{0.1}$)O$_{2-\delta}$  possess different crystal structures at the surface of the pellet (fluorite: \textit{Fm-3m}) and in bulk/powder (Fluorite:\textit{Fm-3m}+Orthorhombic: \textit{Pbcn}). These observations are validated with structural and microstructural studies. The investigation of the origin of this peculiar behavior leads to the design of a new entropy-stabilized fluorite oxide: (Zr$_{0.2}$Hf$_{0.2}$Ce$_{0.2}$Sn$_{0.2}$Mn$_{0.2}$)O$_{2-\delta}$. This is followed by the synthesis of three new entropy-stabilized fluorite oxides: Zr$_{0.2}$Hf$_{0.2}$Ti$_{0.2}$Mn$_{0.2}$Ce$_{0.2}$O$_{2-\delta}$, Zr$_{0.225}$Hf$_{0.225}$Ti$_{0.225}$Mn$_{0.225}$Ce$_{0.1}$O$_{2-\delta}$, and Zr$_{0.2}$Hf$_{0.2}$Ti$_{0.2}$Mn$_{0.2}$Ce$_{0.1}$Ta$_{0.05}$Fe$_{0.05}$O$_{2-\delta}$. X-ray diffraction pattern followed by Rietveld refinement depicts a single-phase fluorite structure on the surface and bulk, which is further validated by scanning electron microscopy and energy-dispersive X-ray spectroscopy analysis. The cyclic heat treatment reveals a reversible transition of multiple phases to single-phase below and above the transition temperature respectively and hence confirms that the thermodynamic stability is dominated by configurational entropy. The thermal conductivity ($\kappa$) in these samples is quite low (1.4--1.7 W·m$^{-1}$·K$^{-1}$), which is explained by the enhanced phonon scattering in high-entropy samples due to high cationic disorder. Impedance spectroscopy measurements show a thermally activated behavior for all the entropy-stabilized samples with their activation energies ranging between 0.3--0.4 eV. These novel fluorite oxides have an optical band gap of 1.6--1.8 eV enabling them to absorb in the visible spectra. Such a lowering of the band gap in these systems as compared to ZrO$_2$, HfO$_2$ or TiO$_2$ based rutile oxides may be ascribed to the presence of high-spin 3d$^5$ Mn$^{2+}$ ions in these entropy-stabilized samples and could make them potential candidates for photocatalysis applications.\\
%

\section*{Conflicts of interest}
There are no conflicts to declare.

\section*{Acknowledgements}
This work was supported by the French Agence Nationale de la Recherche (ANR), through the project NEO (ANR 19-CE30-0030-01). Authors thank Eric Riviere for magnetic measurements.\\

\section{Supporting Information}
XRD patterns of fluorite oxides, Full range XPS of entropy-stabilized oxides, Core-level XPS spectra for Ce-3d and Mn-2p, Thermal diffusivity of novel entropy-stabilized fluorite oxides, optical band-gap, Arrhenius plot for entropy-stabilized oxides, Nyquist and Bode plot and Fitting parameters obtained from the fit3zarcs software for entropy-stabilized oxides.\\
%
%
%

\newpage
\end{document}